\documentclass[aps, showpacs, superscriptaddress, groupedaddress, twocolumn]{revtex4-1} 

\usepackage{amsmath}
\usepackage{amssymb}
\allowdisplaybreaks[2]

\usepackage{graphicx}
\usepackage{subfigure}
\usepackage{longtable} 

\usepackage{multirow}

\usepackage{dcolumn}
\usepackage{bm}
\usepackage{array}
\usepackage{xcolor} 

\begin{document}

\title{Random node reinforcement and $K$-core structure of complex networks}

\author{Rui Ma$^{1,2}$}
\author{Yanqing Hu$^{3}$}
\author{Jin-Hua Zhao$^{1,2,4}$} \email{zhaojh@m.scnu.edu.cn}

\affiliation{
$^1$Guangdong Provincial Key Laboratory of Nuclear Science,
Institute of Quantum Matter,
South China Normal University, Guangzhou 510006, China}

\affiliation{
$^2$Guangdong-Hong Kong Joint Laboratory of Quantum Matter,
Southern Nuclear Science Computing Center,
South China Normal University, Guangzhou 510006, China}

\affiliation{
$^3$Department of Statistics and Data Science, College of Science,
Southern University of Science and Technology, Shenzhen 518055, China}

\affiliation{
$^4$School of Data Science and Engineering,
South China Normal University, Shanwei 516622, China}

\date{\today}

\begin{abstract}
To enhance robustness of complex networked systems, a simple method is introducing reinforced nodes which always function during failure propagation. A random scheme of node reinforcement can be considered as a benchmark for finding an optimal reinforcement solution. Yet there still lacks a systematic evaluation on how node reinforcement affects network structure at a mesoscopic level upon failures.
Here we study this problem through the lens of $K$-cores of networks.
Based on an analytical percolation framework, we first show that, on uncorrelated random graphs, with a critical size of reinforced nodes, an abrupt emergence of $K$-cores is smoothed out to a continuous one, and a detailed phase diagram is derived.
We then show that, with a cost-benefit analysis on random reinforcement, for proper weight factors in cost functions with constant and increasing marginal costs, a gain function shows a unimodality, thus we can analytically find an optimal reinforcement fraction by locating the maximal gain.
In all, our framework offers a gain-oriented analytical perspective to designing robust interconnected systems.
\end{abstract}

\maketitle



\section{Introduction}

Measuring and improving robustness of complex functioning systems against noise and attacks
is vital in complex systems research, both as theoretical and application challenges
\cite{Scheffer-2009}.
This topic becomes increasingly important for real-world systems with intricate interaction patterns,
such as interdependency
\cite{Buldyrev.etal-Nature-2010,
Gao.etal-NatPhys-2012}
and higher-order interactions
\cite{Battiston.etal-NatPhys-2021}
between interacting constituents.
Many external interventions are developed to protect a system from dysfunction or collapse,
such as rewiring links in a system
\cite{Schneider.etal-PNAS-2011},
reinforcing nodes to remove abrupt collapse
\cite{Yuan.Hu.Stanley.Havlin-PNAS-2017,
Xie.etal-Chaos-2019},
introducing redundant interdependency among layers in multiplex networks
\cite{Radicchi.Binaconi-PRX-2017}, and so on.

A typical analytical tool to study robustness of complex systems
is a percolation model on graphs/networks
\cite{Stauffer.Aharony-1994, Newman-2018, Li.etal-PhysRep-2021},
such as site percolation on networks
\cite{Albert.Jeong.Barabasi-Nature-2000, Cohen.etal-PRL-2000, Callaway.etal-PRL-2000}
in which a fraction of nodes are removed and the giant component (GC) of the residual graph
is considered as an indicator of macroscopic connectedness.
While percolation models are usually defined in a random setting, such as a random failure of nodes in networks, the problem of enhancing network robustness intrinsically has an optimization formulation.

In this paper, we study the node reinforcement scheme on networks, a typical measure to improve network robustness.
In an optimization version, the problem can be formulated as finding an optimal scheme of selecting and reinforcing nodes to generate GCs with the largest average size under a given failure model.
While in its random version, the problem can be stated to characterize structural properties of networks under a random node reinforcement scheme, in which a fraction of nodes are randomly selected and reinforced. A random reinforcement scheme can be considered as a benchmark for developing sophisticated optimal ones.
This scheme was adopted to eradicate abrupt transitions in networks with interdependency or group interactions at a macroscopic GC level
\cite{Yuan.Hu.Stanley.Havlin-PNAS-2017, Xie.etal-Chaos-2019}.
Yet how the random scheme affects a network at a level of mesoscopic structure still lacks of study.
Here, we approach this problem through the lens of $K$-cores
\cite{Dorogovtsev.Goltsev.Mendes-PRL-2006, Baxter.etal-PRX-2015}.
Specifically, we study how $K$-cores of networks evolve under random reinforcement scheme upon random node failures.
The $K$-core of a graph is the subgraph after an iterative removal of any node with a degree $< K$,
and its original form and variants act as a structural basis for various processes on networks,
such as a decomposition method to reveal hierarchical and nested structure
\cite{Carmi.etal-PNAS-2007},
structural stability with different interaction patterns
\cite{Zhao.Zhou.Liu-NatCommun-2013, Zhao-JStatMech-2017, Morone.Ferraro.Makse-NatPhys-2019},
and epidemic and behaviourial spreading processes
\cite{Kitsak.etal-NatPhys-2010, Xie.etal-PNAS-2022}.
By combining the original $K$-pruning process with the greedy leaf removal procedure
\cite{Karp.Sipser-IEEE-1981}, a $K$-leaf removal process is defined \cite{AzimiTafreshi.Osat.Dorogovtsev-PRE-2019}. Its significance in network robustness in different structural features and attack settings is further studied
\cite{Shang-PRE-2020, Shang-SIAMJAppMath-2020, Shang-CSF-2021}.

Our main contribution here starts from a model of $K$-core percolation under node reinforcement
and a mean-field theory of the model  in a random scheme on uncorrelated random graphs.
Based on analytical results, we first show a picture different from typical $K$-core percolation,
in which with an increasing fraction of reinforced nodes,
a former hybrid phase transition (a first-order or discontinuous one with a critical singularity)
gradually reduces to a continuous one,
in between there is a mixed case with a new continuous one preceding a hybrid one.
Then, we define a gain function from a cost-benefit perspective for node reinforcement.
By designating an optimal reinforcement fraction to reach the largest gain,
we provide a unified framework to compare schemes of different origins to find an optimal one.

Here is a layout of the paper.
In Sec.\ref{sec:model}, we present our model of $K$-cores on networks with reinforced nodes.
In Sec.\ref{sec:theory}, we develop a mean-field theory for $K$-cores and their GCs on random graphs with randomly distributed reinforced nodes.
In Sec.\ref{sec:result_percolation}, we test our analytical theory on some typical random graph models and real-world networks.
In Sec.\ref{sec:result_cost}, we present a cost-benefit analysis on random node reinforcement. 
In Sec.\ref{sec:conclusion}, we conclude the paper with discussion.

\section{Model}
\label{sec:model}

\begin{figure*}
\begin{center}
 \includegraphics[width = 0.80 \linewidth]{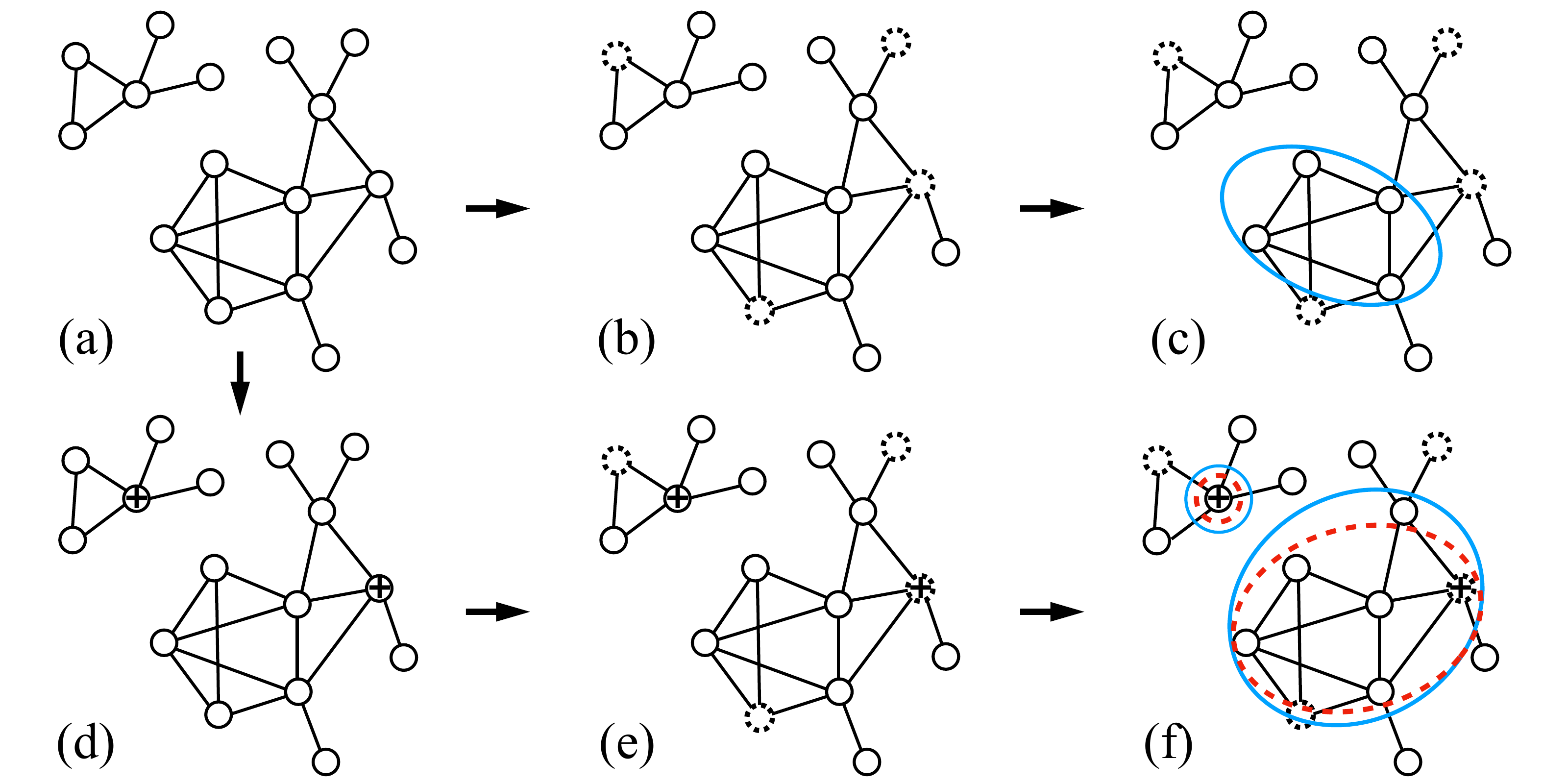} 
\end{center}
\caption{
 \label{fig:model}
Schematics of $K$-cores on an undirected graph with node reinforcement.
(a) A small graph has $16$ nodes and $21$ links, and clusters into two components.
(b) Following the typical $K$-core percolation model, in an initial removal process
$4$ nodes are chosen and removed as indicated in dotted circles.
(c) After $K$-core pruning process on (b), the $2$-core is shown in blue solid circle.
Beware that there is no $3$-core here.
(d) Following our model, starting from the graph in (a), $2$ nodes are selected and reinforced as marked with crosses in circles.
(e) An initial removal of $4$ nodes is applied as the same as in (b).
Beware that a reinforced node is also chosen in the initial removal, yet by definition in our model it cannot be removed.
(f) After $K$-core pruning process in (e), the $2$-core and the $3$-core are enclosed in blue solid and red dashed circles respectively.
Comparing (c) and (f), we can see that, after reinforcing $2$ nodes,
the sizes of $2$-core and $3$-core respectively increase by $3$ and $6$,
and the sizes of their GCs respectively increase by $2$ and $5$.}
\end{figure*}

We consider an undirected graph $G = (V, E)$
with a node set $V$($|V| \equiv N$) and a link set $E$ ($|E| \equiv M$).
For a node $i \in V$, its nearest neighbors constitute a set $\partial i$,
and its cardinality is the degree of $i$ as $k_{i} \equiv | \partial i |$.
The degree distribution $P(k)$ of $G$ is the probability that a randomly chosen node having a degree $k$.
The mean degree of $G$ is simply $c = 2M/N = \sum_{k = 0}^{+\infty} P(k) k$.
A closely related quantity to $P(k)$ common in a percolation theory on graphs
is the excess degree distribution $Q(k)$,
which is the probability that, following a randomly chosen link, an endnode has a degree $k$.
It can be found that $Q(k) = k P(k) / c$.

In the typical $K$-core percolation on a network
\cite{
Dorogovtsev.Goltsev.Mendes-PRL-2006},
the basic step is to remove any node with a degree $< K$ along with all its adjacent links,
leaving the residual subgraph as the $K$-core.
In this paper, we randomly introduce reinforced nodes into a network
\cite{Yuan.Hu.Stanley.Havlin-PNAS-2017}.
A reinforced node functions itself, immune to any removal process,
and further supports its nearest neighbors.
In our percolation model,
initially a fraction $q \in (0,1)$ of nodes are randomly chosen and assigned as being reinforced.
Then an initial removal process follows in which a fraction $1 - p$ ($p \in (0, 1)$ as the initial fraction) of nodes are randomly selected.
The initial removal can be considered as a random failure process, and the ratio $1 - p$ is the strength parameter of failures.
In the initial removal, a selected node can be removed only if it is not reinforced.
On the configuration after reinforcing nodes and the initial removal,
an iterative $K$-core pruning process is carried out.
Beware that, these reinforced nodes cannot be removed in the $K$-core pruning process.
Thus in the residual graph consisting of all functioning nodes,
there are probably reinforced nodes with a degree $< K$.
For the ease of notation, we still name the residual graph as a $K$-core.
Its GC can be easily found with graph search algorithms.
See Fig.\ref{fig:model} for an illustration.

Our model is intrinsically a deterministically irreversible binary-state model on graphs. Like many existing percolation models, with a given configuration of node reinforcement and initial node failures, the final $K$-core structure is independent of order of node removal. This can be proved with a proof by contradiction, which first assumes two distinct final $K$-core configurations, and deduces that there has to be some seed nodes whose different states directly lead to the two distinct configurations, which is actually impossible in deterministically irreversible models.

Here we compare our model with the heterogeneous $K$-core (HKC) model
\cite{Baxter.etal-PRE-2011,
Cellai.etal-PRL-2011,
Cellai.etal-PRE-2013}.
For a reinforced node in our model,
its threshold in $K$-core percolation can be effectively defined as $K = 0$ or $1$ when considering GC of $K$-cores.
In these two models, there are both initial removal process and a $K$-core pruning process
depending on the threshold of each node.
Yet the fundamental difference between them is that, in the HKC model, the initial removal process takes place \textsl{before} randomly assigning thresholds to nodes in a residual graph, while in our model, the initial removal comes in \textsl{after} randomly reinforcing nodes (equivalent to assigning two thresholds, $K$ and $0/1$, randomly to nodes).
With the same parameters of graphs and algorithms,
GCs of $K$-cores in our model are larger than those in the HKC model.
In short, our model has a nontrivial background from network robustness,
and is essentially different from HKC model as a purely extended $K$-core percolation model.

\section{Theory}
\label{sec:theory}

The basic question for our model is to quantitatively estimate sizes of $K$-core and its GC on a graph after random node reinforcement. On uncorrelated random graphs, we can develop a mean-field theory to calculate them in an analytical way.

Our mean-field framework is based on cavity method \cite{Mezard.Montanari-2009}, in which with an assumption of locally tree-like structure of large sparse graphs, we adopt the Bethe-Peierls approximation \cite{Bethe-PRSLA-1935}. This approximation assumes the independence between nodal states of nearest neighbors $\partial i$ of any node $i \in G$ with a prescribed state on a large sparse graph $G$ due to long loops between these neighbors.
In a typical message passing formalism of cavity method for a graphical problem, such as the belief propagation algorithm \cite{Kschischang.Frey.Loeliger-IEEETransInfTheor-2001}, a dimension of $2M$ messages (cavity probabilities) are defined on links of a graph instance and their coupled equations are established. The fixed points of messages after iterations are connected to solutions of the problem. This message passing formalism presents itself often in devising fast algorithms and charactering phase transitions for hard satisfiability and combinatorial optimization problems \cite{Mezard.Montanari-2009}.
Yet in the context of percolation problems on random graphs \cite{Newman.Strogatz.Watts-PRE-2001}, due to simple underlying graphical structure, a much simpler formalism with only $O(1)$ coarse-grained cavity probabilities is possible.

In our percolation model on a graph $G$, two target quantities are fractions of nodes in a $K$-core and those in its GC, denoted as $n$ and $n_{\rm g}$, respectively.
For $G$, we first define two cavity probabilities $\{x, x_{{\rm g}}\} \in [0, 1]$ tailored to dynamical process in our percolation model.
On a randomly chosen link $(i, j) \in G$ between nodes $i$ and $j$, from $i$ to $j$ given that $i$ is in the $K$-core of $G$, we define $x$ and $x_{{\rm g}}$, respectively, as the probability that $j$ is in $K$-core and in its GC. Under the Bethe-Peierls approximation, we derive self-consistent equations for $x$ and $x_{\rm g}$. With their stable fixed solutions, we finally calculate $n$ and $n_{\rm g}$.
All the relevant equations are listed as below.
\begin{widetext}
\begin{eqnarray}
x
\label{eq:x}
&& = q + (1 - q) p \sum _{k = K}^{+\infty} Q(k) \sum _{s = K - 1}^{k - 1}
{k - 1 \choose s}
x^{s} (1 - x) ^{k - 1 - s}, \\
x_{{\rm g}}
\label{eq:xg}
&& = q \sum _{k = 1}^{+ \infty} Q(k) \sum _{s = 1}^{k - 1} 
{k - 1 \choose s}
[x^{s} - (x - x_{\rm g})^{s}]
(1 - x)^{k - 1 - s}  \nonumber \\
&&
+ (1 - q) p \sum _{k = K}^{+ \infty} Q(k) \sum _{s = K - 1}^{k - 1} 
{k - 1 \choose s}
[x^{s} - (x - x_{\rm g})^{s}]
(1 - x)^{k - 1 - s},\\
n
\label{eq:n}
&& = q + (1 - q) p \sum _{k = K}^{+ \infty} P(k) \sum _{s = K}^{k} 
{k \choose s}
x^{s} (1 - x)^{k - s},\\
n_{{\rm g}}
\label{eq:ng}
&& = q \sum _{k = 1}^{+ \infty} P(k) \sum _{s = 1}^{k} 
{k \choose s}
[x^{s} - (x - x_{\rm g})^{s}]
(1 - x)^{k - s} \nonumber \\
&& + (1 - q) p \sum _{k = K}^{+ \infty} P(k) \sum _{s = K}^{k} 
{k \choose s}
[x^{s} - (x - x_{\rm g})^{s}]
(1 - x)^{k - s}.
\end{eqnarray}
\end{widetext}

We first briefly explain Eqs.(\ref{eq:x}) and (\ref{eq:xg}).
We follow the setting in defining $x$ and $x_{\rm g}$ with a randomly chosen link $(i, j) \in G$ from $i$ to $j$.
For Eq.(\ref{eq:x}), if $j$ is in $K$-core, there are two possibilities.
First, $j$ is a reinforced node, then it is surely in $K$-core. Thus we have the first term on right-hand side (RHS) of Eq.(\ref{eq:x}).
Second, $j$ is not a reinforced node, and survives the initial removal. For $j$ to be further in $K$-core, it must have at least $K - 1$ nodes among its nearest neighbors also in $K$-core besides $i$, since $i$ is already in $K$-core. Thus we have the second term on RHS of Eq.(\ref{eq:x}).
For Eq.(\ref{eq:xg}), if $j$ is in GC of $K$-core, it must first be in $K$-core.
Correspondingly, there are also two possibilities.
First, $j$ is a reinforced node. For $j$ to be further in GC of $K$-core,
it must have at least one nearest neighbor which are in GC of $K$-core besides $i$,
since by definition we don't specify that $i$ is in GC of $K$-core or not.
Thus we have the first term on RHS of Eq.(\ref{eq:xg}).
Second, $j$ is not a reinforced node, and remains after the initial removal.
For $j$ to first in $K$-core and then in its GC, it must have at least $K - 1$ nearest neighbors also in $K$-core  besides $i$ since $i$ is already in $K$-core, among which there are at least one neighbor in GC of $K$-core. Thus we have the second term on RHS of Eq.(\ref{eq:xg}).

In Eqs.(\ref{eq:n}) and (\ref{eq:ng}), we estimate the possibility of a randomly chosen node $i \in G$ to be in $K$-core and in GC of $K$-core respectively, which follows a similar logic in Eqs.(\ref{eq:x}) and (\ref{eq:xg}). A major difference in their probabilistic forms is that, in formulating Eqs.(\ref{eq:x}) and (\ref{eq:xg}), we follow a randomly chosen link $(i, j)$ and consider the probability of $j$'s state. Thus in establishing them we adopt the excess degree distribution $Q(k)$. Yet in Eqs.(\ref{eq:n}) and (\ref{eq:ng}), we calculate the probability of state of a randomly chosen node in $G$, thus we adopt the degree distribution $P(k)$.

In Appendix, we modify those summation terms on degrees in Eqs.(\ref{eq:x}) - (\ref{eq:ng}) into a form with generating functions of degree distributions, which are more explicit in performing numerical calculation.

%
\begin{figure*}
\begin{center}
 \includegraphics[width = 0.99 \linewidth]{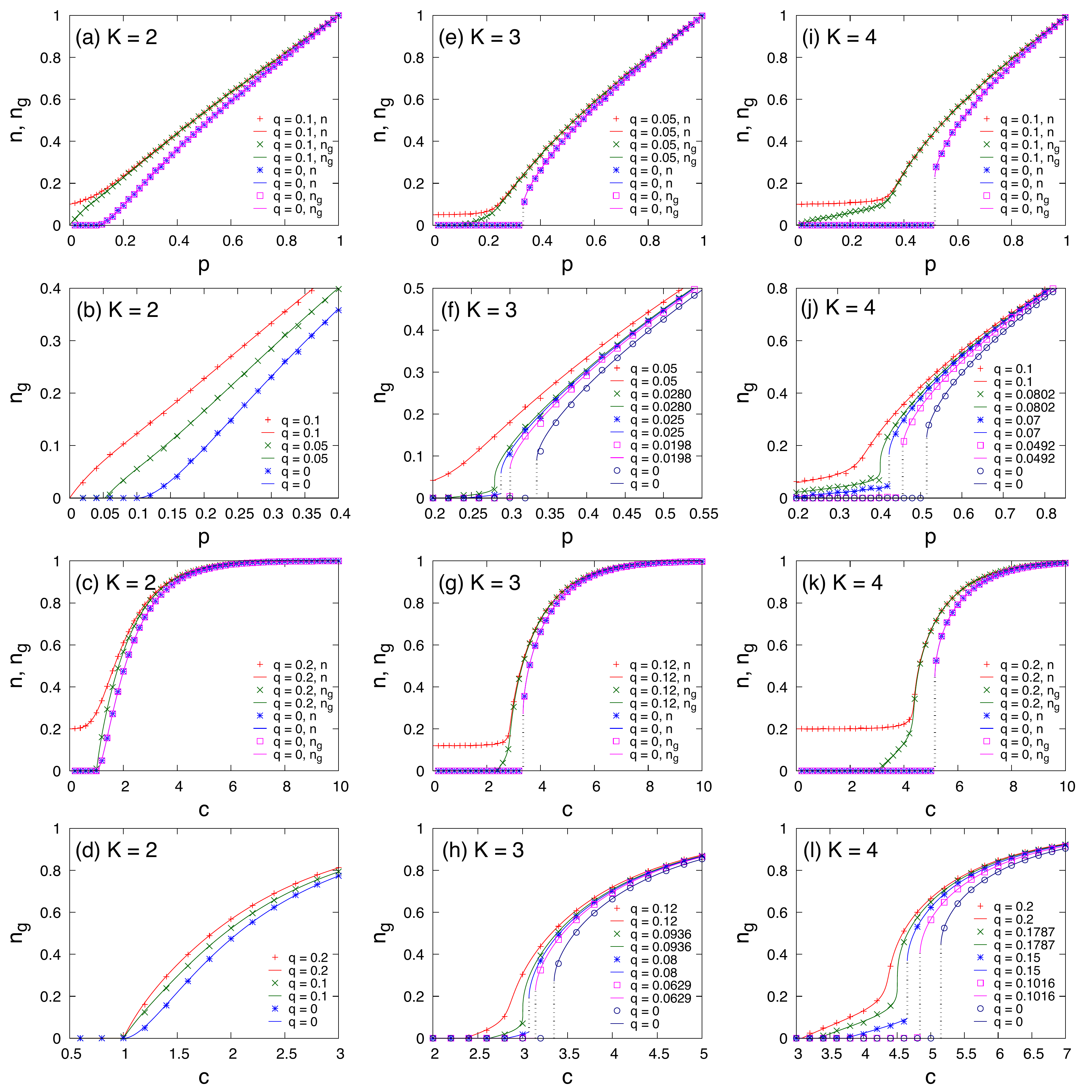}
\end{center}
\caption{
 \label{fig:er}
$K$-cores on ER random graphs with random node reinforcement.
(a)-(b) Fractions of $K$-core $n$ and its GC $n_{\rm g}$ on ER random graphs with $c = 10$ for $K = 2$.
(c)-(d) $n$ and $n_{\rm g}$ on ER random graphs with $p = 1$ for $K = 2$.
(e)-(h) and (i)-(l) have a similar format with (a)-(d) yet are for $K = 3$ and $4$, respectively.
In (f), (h), (j), and (l),
$q$ are chosen with values in the five cases of
$q = 0, q \approx q^{t}, q^{t} < q < q^{c}, q \approx q^{c}, q^{c} < q$.
See in the main text.
Each sign is a result from simulation on a graph instance with a node size $N = 10^5$.
Solid lines are results from analytical theory on infinitely large graphs,
and dotted lines denote jumps in analytical results.}
\end{figure*}
\begin{figure*}
\begin{center}
 \includegraphics[width = 0.99 \linewidth]{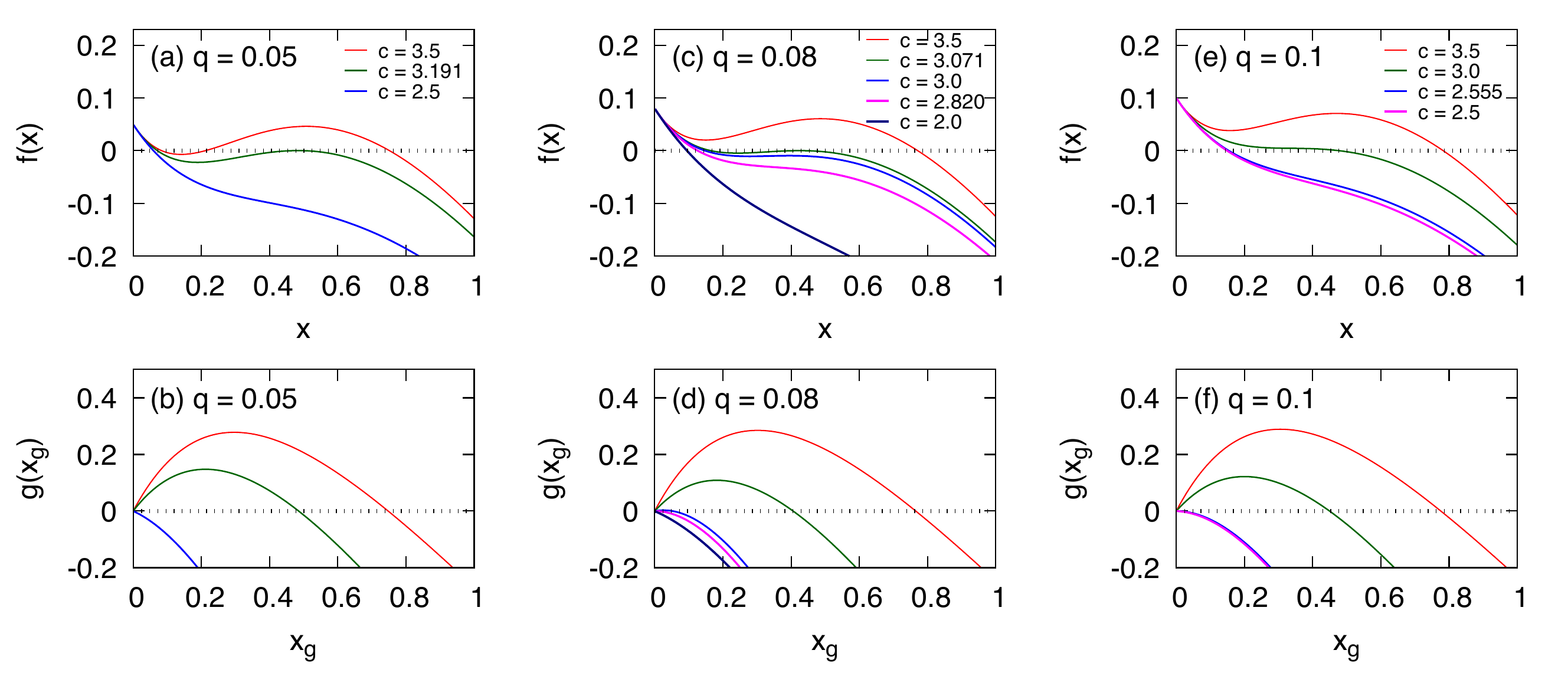}
\end{center}
\caption{
 \label{fig:er_fx}
Fixed point analysis of self-consistent equations on ER random graphs.
We have $K = 3$ and $p = 1.0$.
In (a)-(b), (c)-(d), and (e)-(f), we consider $q = 0.05, 0.08, 0.1$, respectively,
in which $f(x)$ is shown in an upper subfigure, and $g(x_{\rm g})$ is shown in a lower subfigure with corresponding stable $x$.
In the legend, $c$ with four significant digits are thresholds of continuous or hybrid transitions depending on the scenario of fixed point behaviors with corresponding $q$.
Lines are results from analytical theory on infinitely large graphs.}
\end{figure*}
\begin{figure*}
\begin{center}
 \includegraphics[width = 0.99 \linewidth]{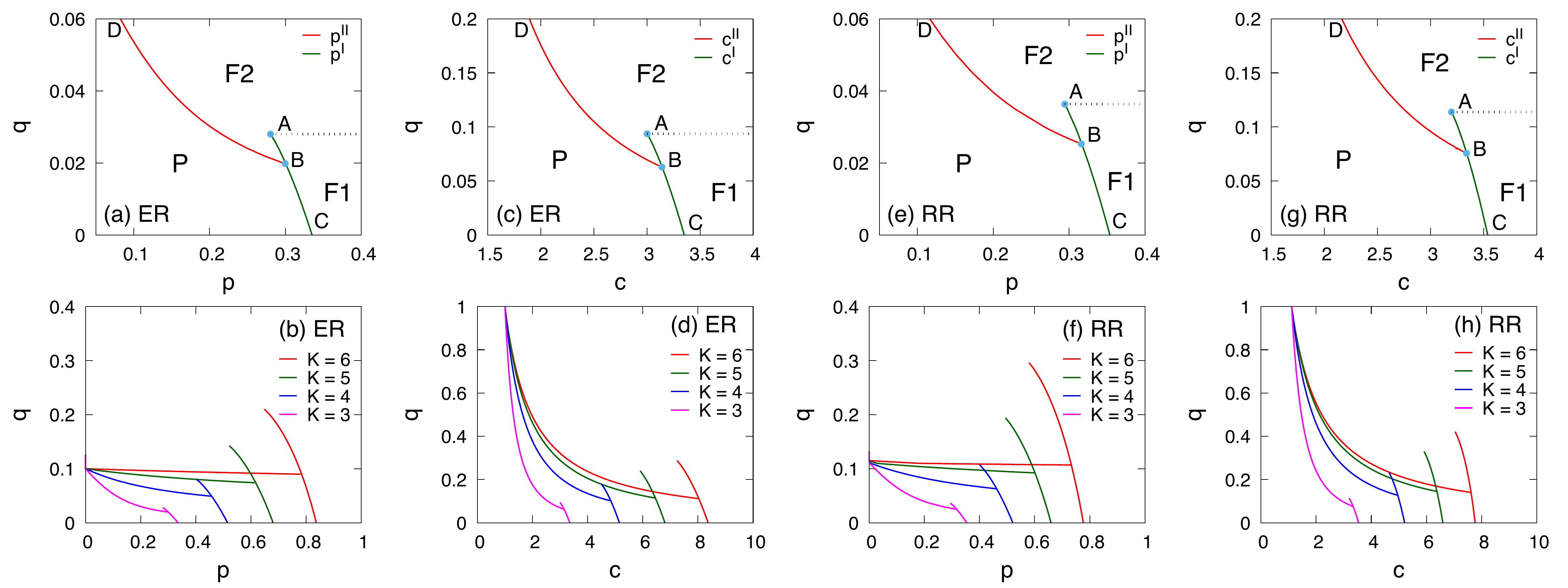}
\end{center}
\caption{
 \label{fig:regimes}
Phase diagrams from our analytical framework on ER random and RR graphs with an infinitely large size.
(a) Phase diagram on ER random graphs with $c = 10$ for $K = 3$.
Line segment BD denotes continuous transition points of $K$-cores,
AB hybrid transition points preceded by continuous ones,
and BC proper hybrid transition points with jumps starting from $0$.
P, F1, and F2 respectively denote phases in which
there is no percolation, percolation after a hybrid transition,
and percolation after a second-order transition.
(b) Same with (a) yet with $K = 3, 4, 5, 6$ on a larger scale.
(c)-(d) Similar with (a)-(b) for phase diagrams on ER random graphs with $p = 1$ on varying $c$.
(e)-(f) Similar with (a)-(b) for phase diagrams on RR graphs with $k_{0} = 10$.
(g)-(h) Similar with (c)-(d) for phase diagrams on diluted RR graphs with $k_{0} = 10$ and $p = 1$.}
\end{figure*}
\begin{figure*}
\begin{center}
 \includegraphics[width = 0.99 \linewidth]{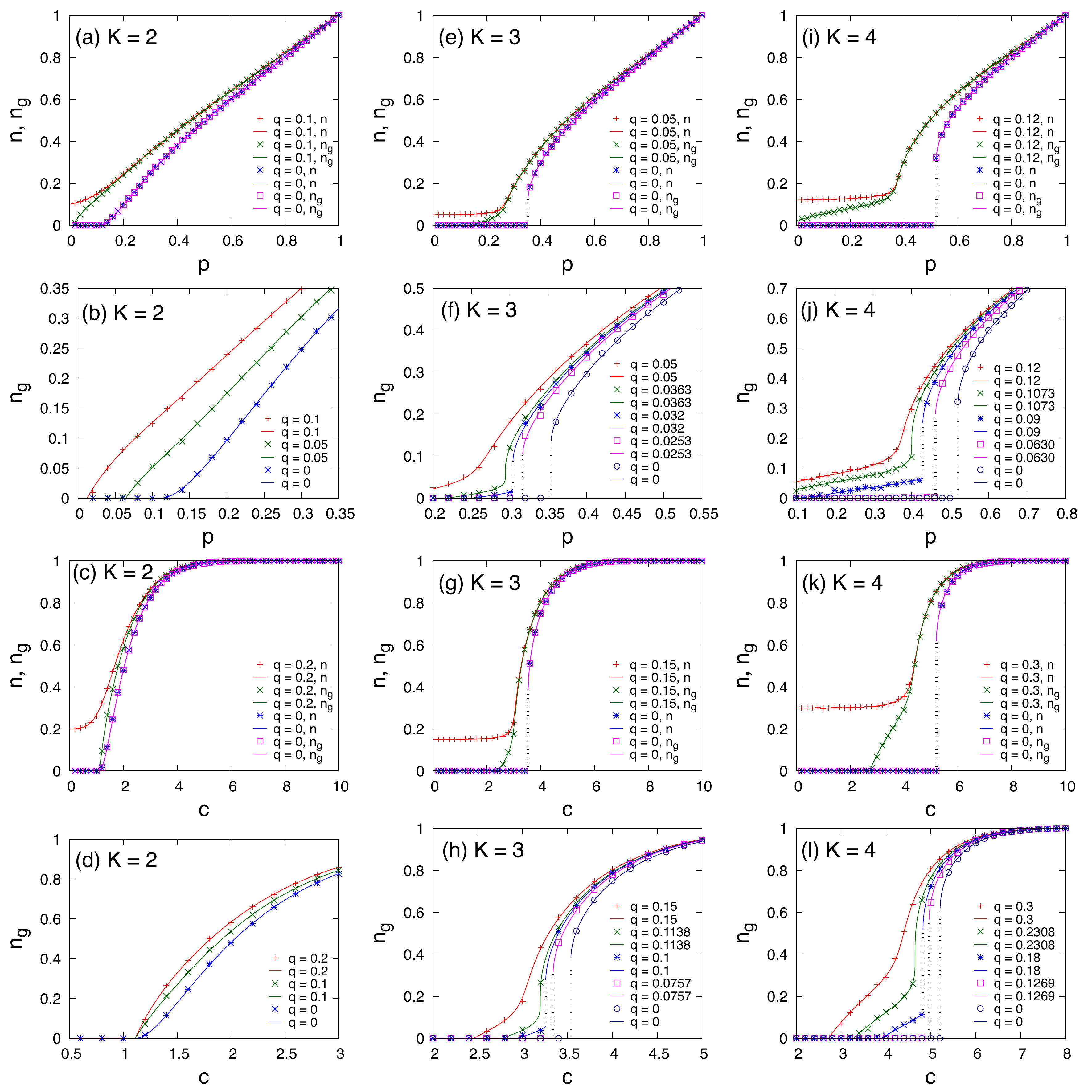}
\end{center}
\caption{
 \label{fig:rr}
$K$-cores on RR graphs with random node reinforcement.
(a)-(b) Fractions of $K$-core $n$ and its GC $n_{\rm g}$ on RR graphs with $k_{0} = 10$ for $K = 2$.
(c)-(d) $n$ and $n_{\rm g}$ on diluted RR graphs with $k_{0} = 10$ and $p = 1$ for $K = 2$.
(e)-(h) and (i)-(l) have a similar format with (a)-(d) yet are for $K = 3$ and $4$, respectively.
In (f), (h), (j), and (l),
$q$ are chosen with values in the five cases of
$q = 0, q \approx q^{t}, q^{t} < q < q^{c}, q \approx q^{c}, q^{c} < q$.
See in the main text.
Each sign is a result from simulation on a graph instance with a node size $N = 10^5$.
Solid lines are results from analytical theory on infinitely large graphs,
and dotted lines denote jumps in analytical results.}
\end{figure*}
\begin{figure*}
\begin{center}
 \includegraphics[width = 0.99 \linewidth]{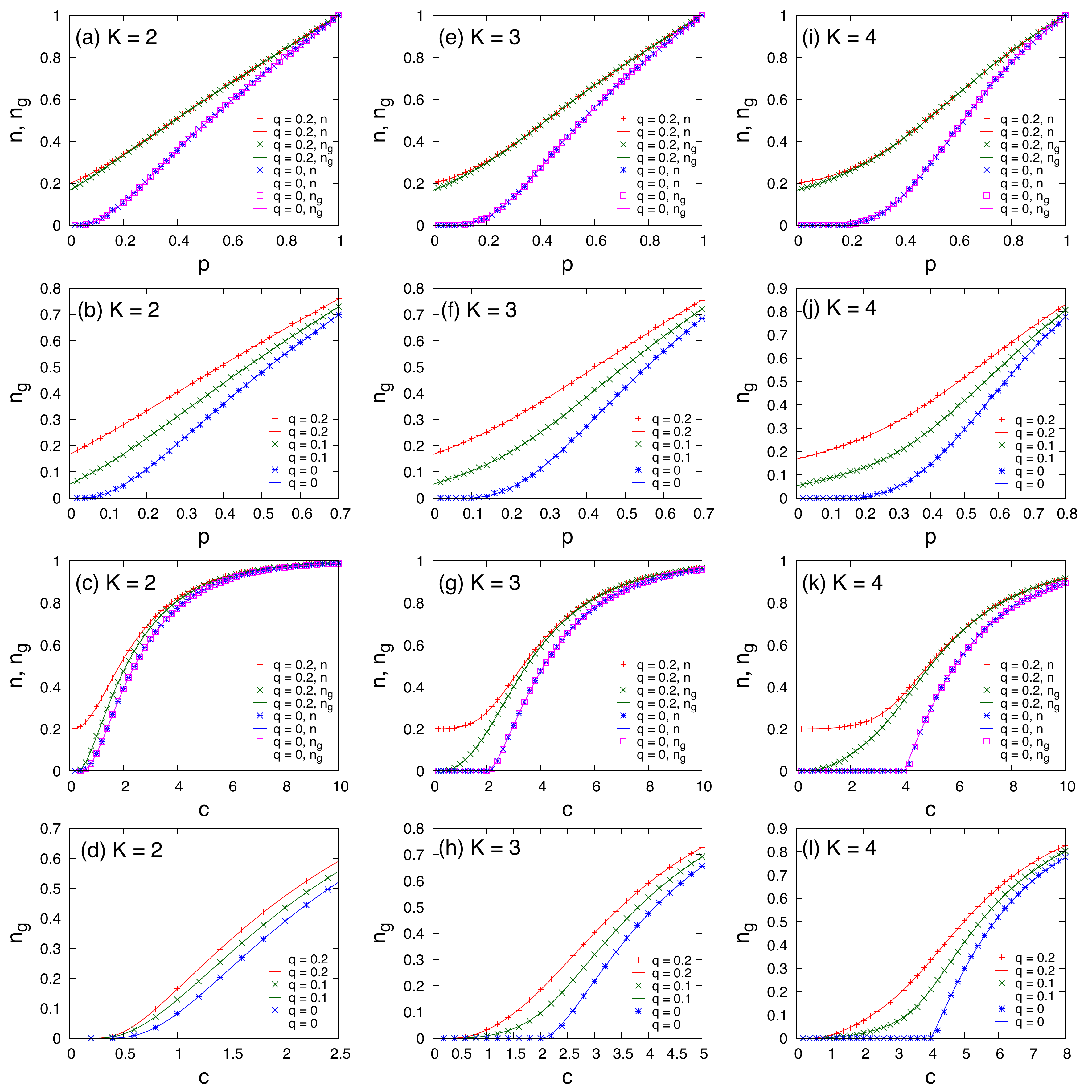}
\end{center}
\caption{
 \label{fig:sf}
$K$-cores on SF networks with random node reinforcement.
(a)-(b) Fractions of $K$-core $n$ and its GC $n_{\rm g}$
on a SF network instance generated with configuration model
with a degree exponent $\gamma = 2.5$ and a node size $N = 10^5$ for $K = 2$.
In the graph construction, we have $k_{{\rm min}} = 6$ and $k_{{\rm max}} = \sqrt{N}$.
Each sign is a result from simulation on the given graph instance with a node size $N = 10^5$.
Lines are results from analytical theory based on the empirical degree distribution of the graph instance.
(c)-(d) $n$ and $n_{\rm g}$ on asymptotical SF networks generated with static model with $\gamma = 3.0$ for $K = 2$.
Each sign is a result from simulation on a graph instance with a node size $N = 10^5$.
Lines are results from analytical theory on infinitely large graphs.
(e)-(h) and (i)-(l) have a similar format with (a)-(d) yet for $K = 3$ and $4$, respectively.}
\end{figure*}
\begin{figure*}
\begin{center}
 \includegraphics[width = 0.99 \linewidth]{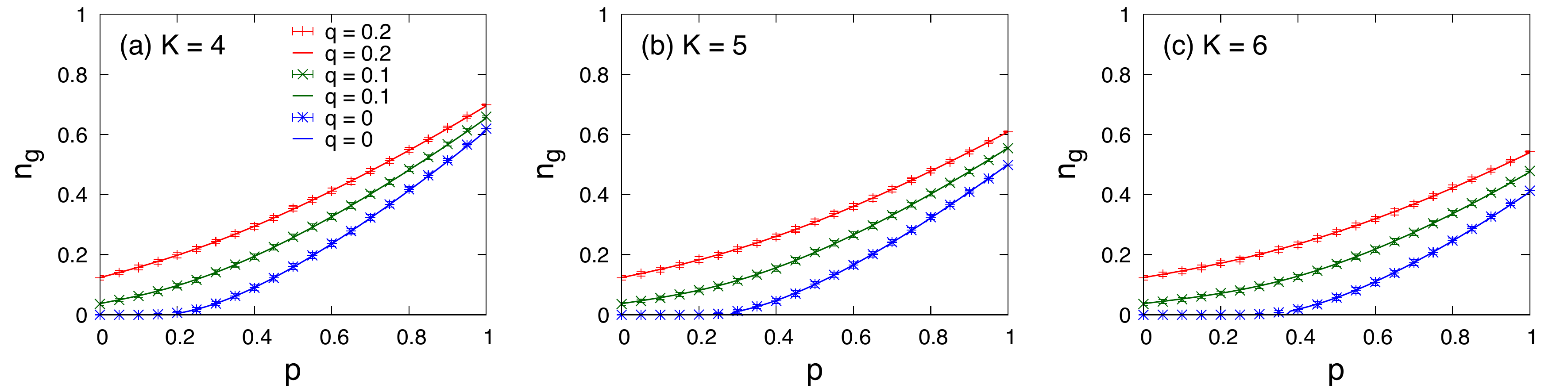}
\end{center}
\caption{
 \label{fig:real}
$K$-cores on a protein interaction network with random node reinforcement.
In (a)-(c), we consider the cases of $K = 4, 5, 6$, respectively.
Signs are average results with standard deviations for $n_{g}$ from simulation of our percolation model on the network instance. For each simulation result of $n_{g}$, we generate $100$ different configurations of node reinforcement and initial removal to calculate their average and standard deviation. Solid lines are for the analytical prediction of $n_{g}$ based on our mean-field framework.}
\end{figure*}
\begin{figure*}
\begin{center}
 \includegraphics[width = 0.99 \linewidth]{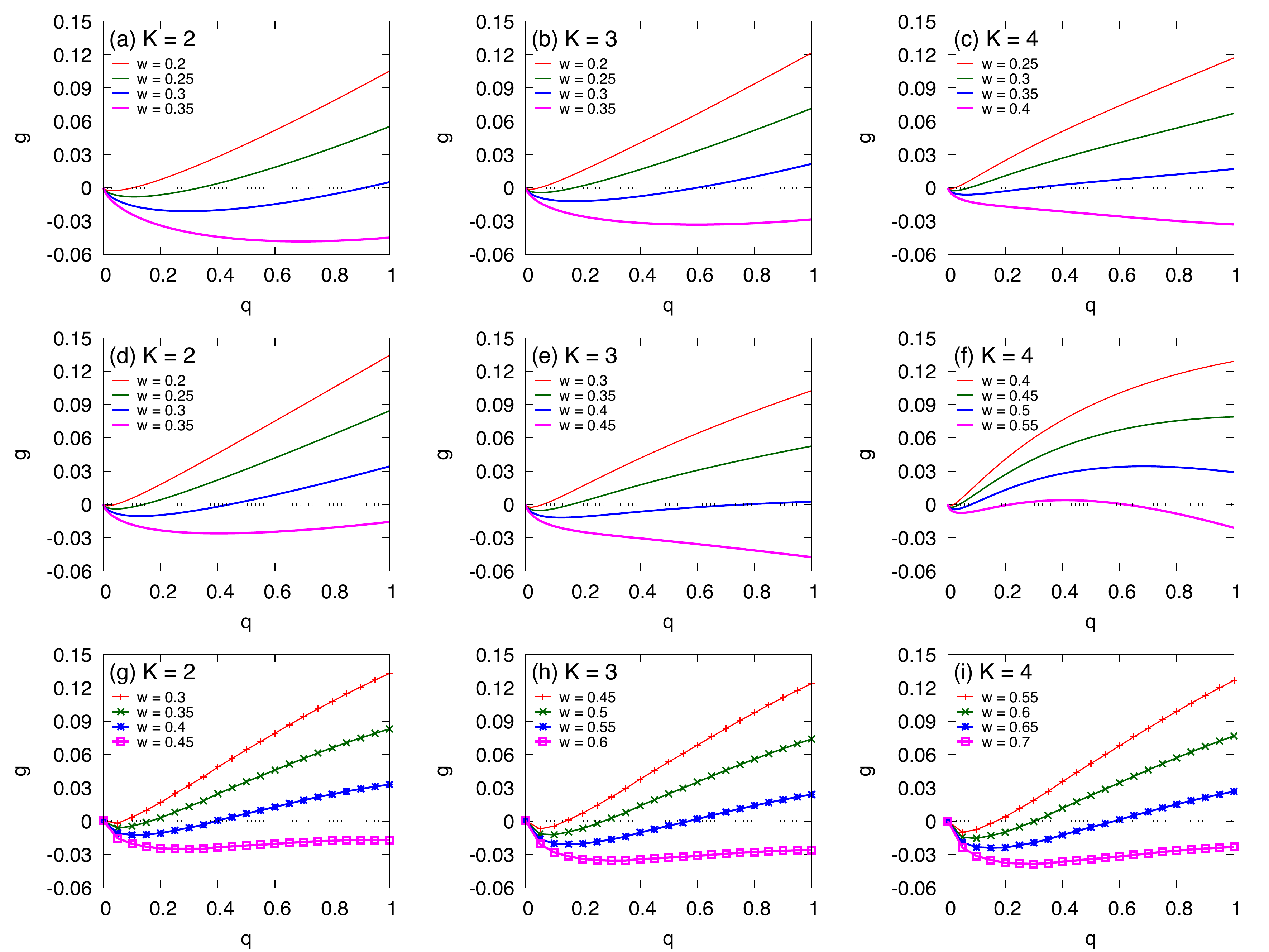}
\end{center}
\caption{
 \label{fig:gain_sublinear}
Gain function on random graphs and a real-world network.
The initial removal fraction is $p = 0.7$.
In the cost term, $\alpha = 0.8$.
Results of our model with $K = 2, 3, 4$ are listed in columns from left to right, respectively.
(a-c) Gain function on ER random graphs with $c = 10$.
Solid line are results from mean-field theory on infinitely large graphs.
(d-f) Gain function on SF networks generated by static model with $\gamma = 3.0$ and $c = 10$.
Solid lines are also results from mean-field theory on infinitely large graphs.
(g-i) Gain function on a protein interaction network with a mean degree $c \approx 10.3$.
Each sign is a simulation result averaged on $1000$ realizations of node reinforcement and initial removal configurations.}
\end{figure*}
\begin{figure*}
\begin{center}
 \includegraphics[width = 0.99 \linewidth]{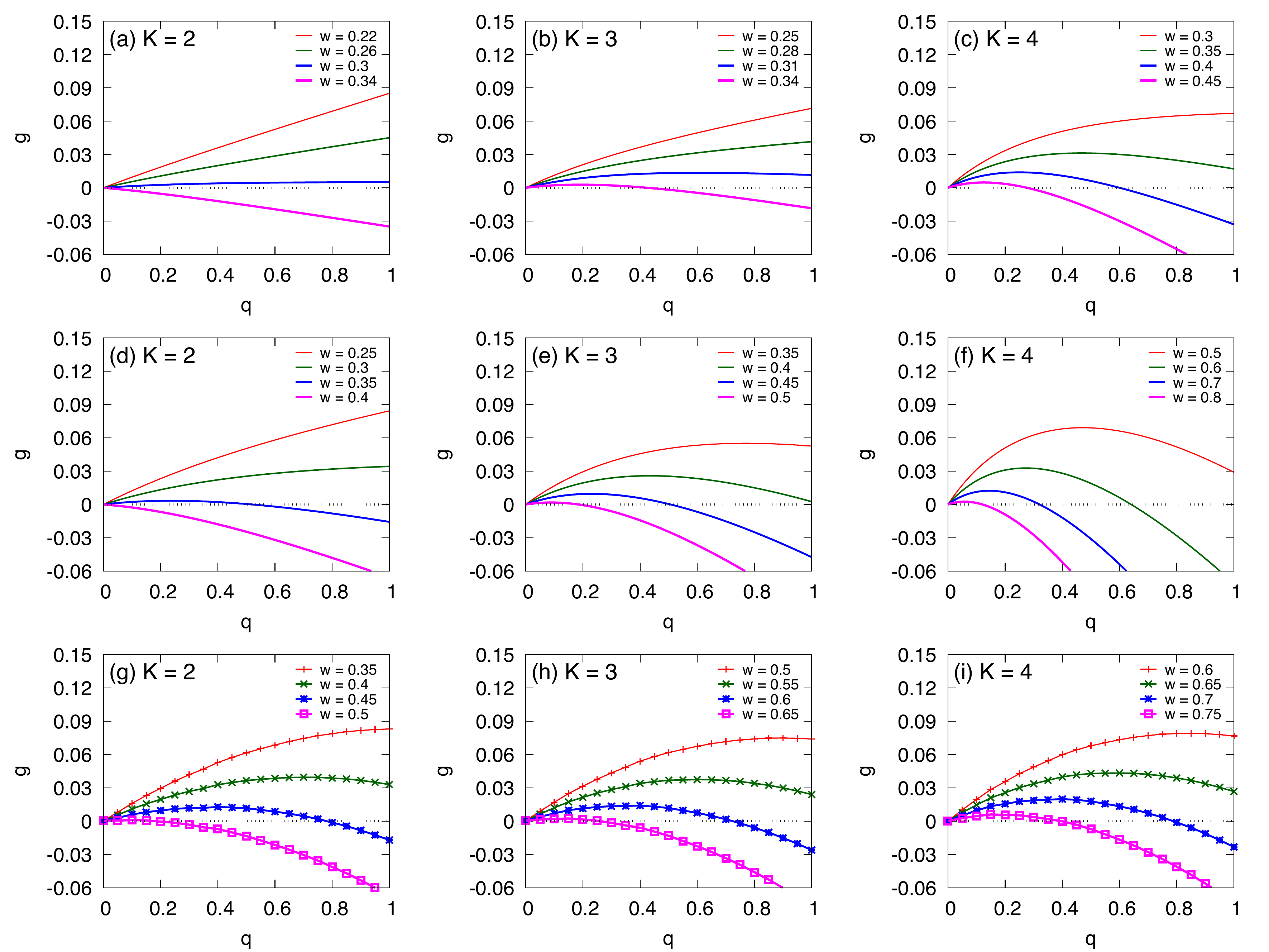}
\end{center}
\caption{
 \label{fig:gain_linear}
Same with Fig.\ref{fig:gain_sublinear}, yet in the cost term $\alpha = 1$.}
\end{figure*}
\begin{figure*}
\begin{center}
 \includegraphics[width = 0.99 \linewidth]{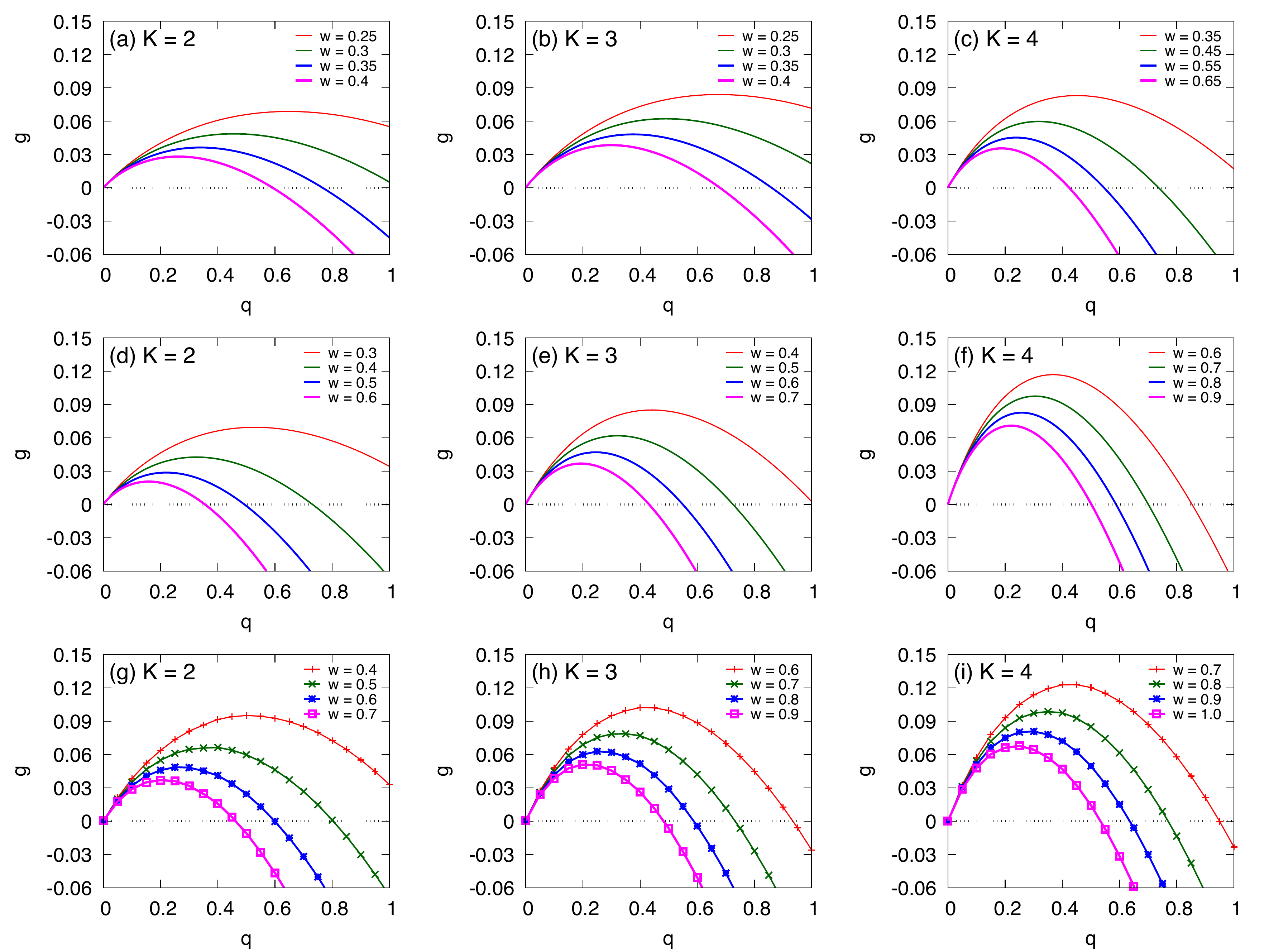}
\end{center}
\caption{
 \label{fig:gain_superlinear}
Same with Fig.\ref{fig:gain_sublinear}, yet in the cost term $\alpha = 1.5$.}
\end{figure*}

\section{Result: A percolation analysis}
\label{sec:result_percolation}

We test our mean-field theory of percolation model on some typical random graph models.
First we consider Erd\"{o}s-R\'{e}nyi (ER) random graphs
\cite{Erdos.Renyi-PublMath-1959, Erdos.Renyi-Hungary-1960}.
For ER random graphs with a mean degree $c$, we have a Poisson degree distribution as
\begin{eqnarray}
P(k) = {\rm e}^{- c} \frac {c^k}{k!}.
\end{eqnarray}
%
Results of $K$-core sizes are shown in Fig.\ref{fig:er}.
We can see that theoretical and simulation results agree very well.
The general effect of reinforced nodes on $K$-core formation is that,
they act as seeds of disconnected functioning clusters.
Only when the mean degree $c$ of an original graph and a reinforcement fraction $q$ are large enough,
those disconnected components can merge into a macroscopic one.
There are three points in the general picture of these results we would emphasize.
The first point is shown in the first and third rows in Fig.\ref{fig:er}.
We show that,
with a nontrivial $q$ a significant difference between $n$ and $n_{\rm g}$ exists
when a mean degree $c$ is relatively small,
while in the original $K$-core percolation on ER random graphs,
there is only an indiscernible difference between them.
The second point is shown in the second and fourth rows in Fig.\ref{fig:er}.
We show that $q$ pushes up $n_{\rm g}$ especially significantly around transition points,
and moves the birth point of a $K$-core to a smaller $p$ or $c$.
The third point is specifically for those hybrid transitions in the cases of $K \geqslant 3$.
A detailed analysis shows that with an increasing $q$ from zero,
$n_{\rm g}$ consecutively shows a hybrid transition,
a two-stage process in which a new continuous one precedes a hybrid one,
and finally a fully continuous one.
Correspondingly, we denote two critical values for $q$:
$q_{t}$ as the $q$ at which a new continuous transition first sets in, and
$q_{c}$ as the $q$ at which a hybrid transition recedes into a continuous one.
This scenario of smoothing an abrupt transition into a continuous one by tuning a control parameter
also shows in other percolation problems,
such as introducing reinforced nodes into multilayer networks
\cite{Yuan.Hu.Stanley.Havlin-PNAS-2017}
or single networks with group dependency
\cite{Xie.etal-Chaos-2019},
a percolation model with interactions among next-nearest neighbors
on single networks with both unidirectional and undirected links
\cite{Xie.etal-PNAS-2022},
interdependent networks with variable coupling strengths
\cite{Parshani.Buldyrev.Havlin-PRL-2010},
single networks with both dependency and connectivity links
\cite{Parshani.etal-PNAS-2011},
and weak percolation on multiplex networks with overlapping edges \cite{Baxter.etal-CSF-2022}.
An intuitive explanation for this scenario is that, all these models are driven by two competing adversary forces with different transition patterns. For example, in our model, the two forces are the $K$-core pruning process which leads to a shrinkage of functioning components, and the node reinforcement which connects separate functioning components to form a larger one. The former behaves in a discontinuous way with $K \geqslant 3$ on uncorrelated random graphs, yet the latter works in a continuous manner when it follows a random scheme. By tuning a relative strength parameter between two driving forces, as the reinforcement fraction $q$ in our model, a smooth crossover between hybrid and continuous transitions underlying the two forces can be realized.

Here we elucidate the phase transition behavior through a fixed point analysis.
We denote $F(x;q)$ and $G(x_{\rm g}, x;q)$ respectively as the right-hand side of Eqs.(\ref{eq:x}) and (\ref{eq:xg}), and define $f(x) \equiv - x + F(x;q)$ and $g(x_{\rm g}) \equiv -x_{\rm g} + G(x_{\rm g}, x; q)$. From $f(x) = 0$ and $g(x_{\rm g}) = 0$, we can read fixed $x$ and $x_{\rm g}$. In Fig.\ref{fig:er_fx},
we show the fixed points of $x$ and $x_{\rm g}$ on ER random graphs for $K = 3$ and $p = 1.0$.
We test some representative $q$ in cases of $[0, q_{t}), (q_{t}, q_{c}), (q_{c}, 1]$, respectively.
From the behavior of fixed points we ascertain orders and thresholds of transitions.
Here we only explain the case in $q = 0.08 \in (q_{t}, q_{c})$ in Figs.\ref{fig:er_fx} (c) and (d).
At $c < 2.820$, there is only one fixed solution for $x$ and the corresponding stable $x_{\rm g} = 0$.
At $2.820 < c < 3.071$, there is also only one fixed solution for $x$, yet a second (and stable) fixed solution of $x_{\rm g}$ emerges continuously, leading to a small yet nontrivial $n_{\rm g}$.
At $c \approx 3.071$,
a second (and stable) fixed solution of $x$ shows after a jump from $x \approx 0.164$ to $x \approx 0.424$, further leads to sudden jumps in $x_{\rm g}$, $n$, and $n_{\rm g}$. See the bulge in formation in $f(x)$ when $x > 0.2$.
At $c > 3.071$, the stable $x$ and $x_{\rm g}$ increase continuously thereafter.
We can see that, a jump in $K$-core fractions here is mainly driven by the abrupt behavior of stable solution of $x$, and a continuous transition is mainly driven by the instability of the trivial fixed point $x_{\rm g} = 0$.

For the above physical picture of phase transitions on general random graphs,
there are conditions to calculate their critical values.
In the case of $q \leqslant q_{c}$,
we can calculate the hybrid phase transition point $(c^{\rm I}, x^{\rm I})$ or $(p^{\rm I}, x^{\rm I})$.
Take $(c^{\rm I}, x^{\rm I})$ for an example,
it satisfies the condition as:
\begin{eqnarray}
\frac {\partial F(x; q)}{\partial x} \bigg|_{c = c^{\rm I}, x = x^{\rm I}} = 1.
\end{eqnarray}
To further calculate $q_{c}$, we have
\begin{eqnarray}
\frac {{\rm d} c^{\rm I}}{{\rm d} x^{\rm I}} \bigg| _{q = q_{c}} = 0.
\end{eqnarray}
In the case of $q \geqslant q_{t}$,
we can calculate the continuous phase transition point $(c^{\rm II}, x^{\rm II})$ or $(p^{\rm II}, x^{\rm II})$.
Take $(c^{\rm II}, x^{\rm II})$ for an example,
it satisfies the condition as:
\begin{eqnarray}
\frac {\partial G(x_{\rm g},x;q)}{\partial x_{\rm g}} \bigg|_{c = c^{\rm II}, x = x^{\rm II}, x_{\rm g} = 0} = 1.
\end{eqnarray}
To further calculate $q_{t}$, we have
\begin{eqnarray}
\frac {{\rm d} c^{\rm II}}{{\rm d}x^{\rm II}} \bigg|_{q = q_{t}} = 0.
\end{eqnarray}
The above transition points and the critical values of $q$
can be calculated by these conditions together with Eqs.(\ref{eq:x}) and (\ref{eq:xg}).

We further clarify the hybrid phase transition behavior in our model with phase diagrams.
With an increasing $q$, we list and connect all its critical control parameters.
In Fig.\ref{fig:regimes}, we show three branches of critical parameters denoting second-order transitions, hybrid ones, and hybrid ones preceded by continuous ones. These branches act as boundaries between three phases, which are a non-percolated phase with $n_{\rm g} = 0$ (P), a percolated phase with $n_{\rm g} > 0$ after hybrid transitions (F1), and a percolated phase with $n_{\rm g} > 0$ after second-order transitions (F2). It is easy to see that, points A correspond to the largest $q$ on hybrid transition lines, equivalently $q_{c}$, and points B correspond to the smallest $q$ on second-order transition lines, equivalently $q_{t}$.

We then consider our analytical framework on regular random (RR) graphs.
A RR graph has a uniform degree distribution
$P(k) = \delta (k, k_{0})$ with $k_{0} \geqslant 2$.
To generate a graph instance with a heterogeneous degree profile,
we dilute a RR graph by randomly removing a fraction $1 - \rho$ $\in (0, 1)$ of its links.
For a diluted graph instance, the mean degree is $c = \rho k_{0}$ and the degree distribution is
\begin{eqnarray}
P(k) = {k_{0} \choose k} \rho ^k (1 - \rho) ^{k_{0} - k}.
\end{eqnarray}
Results of $K$-core sizes are shown in Fig.\ref{fig:rr}, and a phase diagram is shown in Fig.\ref{fig:regimes}.
Their physical picture is qualitatively similar with that on ER random graphs.

We then consider scale-free (SF) networks
\cite{Barabasi.Albert-Science-1999}
which follow a power-law degree distribution as $P(k) \propto k^{- \gamma}$ with $\gamma$ as a degree exponent.
The SF property is ubiquitous in real-world datasets as a partial result of rich dynamics in network formation processes.
Two typical ways to construct SF networks are the configuration model
\cite{Newman.Strogatz.Watts-PRE-2001}
and the static model
\cite{Goh.Kahng.Kim-PRL-2001,
Catanzaro.PastorSatorras-EPJB-2005}.
Both models can generate SF networks with a wide range of degree exponents as $\gamma > 2.0$. They intrinsically represent two complementary approaches to SF networks: the former empirically reflects SF property of degree profiles yet only applies to a finite graph size, while the latter approximates SF property of degrees asymptotically yet naturally offers an analytical way to tackle degree property on infinitely large SF networks.

First we consider the configuration model,
which can approximately generate a graph instance with any proper degree distribution.
This model first generates a degree sequence based on the given degree distribution,
then assigns nodes in a null graph with degrees sampled from the degree sequence,
and finally establishes proper links among nodes to constructs a network instance.
The key parameters for SF networks in this model are $\gamma$ as the degree exponent, $N$ the node size, $k_{\rm{min}}$ the minimal degree of nodes, and $k_{\rm{max}}$ the maximal degree of nodes.
For simplicity, we set $k_{\rm{max}} = \sqrt N$ to remove degree-degree correlation in networks. For an analytical result on SF network instances generated with this model, we simply read a SF network instance, count its empirical degree distribution, and feed it into our theoretical equations.

We then consider the static model to construct asymptotical SF networks.
This model is generally a random process of establishing links between node pairs
with probabilities proportional to their node weights following a power-law distribution.
To construct a SF network with a degree exponent $\gamma$
and a node size $N$ of a set $V = \{1, 2, ..., N\}$, the weight of a node $i \in V$
is $w_{i} \propto i ^{- \xi}$ with an intermediary parameter $\xi  \equiv 1 / (\gamma - 1)$.
For $\gamma \geqslant 3.0$,
the degree-degree correlation in those generated networks is negligible.
Static model, unlike configuration model,
admits an explicitly analytical form of degree distribution for large graph instances.
The degree distribution of SF networks from static model is
\begin{eqnarray}
P(k) = \frac {1}{\xi}\frac {[c (1 - \xi)]^k}{ k!} {\rm E}_{- k + 1 + \frac{1}{\xi}} [c (1 - \xi)].
\end{eqnarray}
The special function ${\rm E}_{a}(x)$ is a general exponential integral function as
${\rm E}_{a}(x) \equiv \int _{1}^{\infty} \mathrm{d}t e^{-xt} t^{- a}$
with $a, x > 0$. For large $k$, we have $P (k) \propto k^{- \gamma}$.

Results on SF networks with the two generation models are shown in Fig.\ref{fig:sf}. Its physical picture is quite similar to that in ER random and RR graphs with $K = 2$, since there is no abrupt emergence of $K$-cores due to the statistical property of power-law degree distributions of SF networks
\cite{Dorogovtsev.Goltsev.Mendes-PRL-2006}.

We further test our model on a real-world network instance
\cite{Vinayagam.etal-ScienceSignaling-2011}.
The original protein interaction network is intrinsically a directed one involving both directional links
(an ordered node pair as a node pointing to the other) and multi-edges (more than one directional links between an ordered node pair). To extract the underlying interaction topology of the network, we ignore directions of links and merge multi-edges between node pairs. Finally, the original network dataset reduces to an undirected network with a node size $N = 6339$ and a link size $M = 32706$,  equivalently $c \approx 10.3$.
To make an analytical prediction for GC of $K$-core on a real-world network, just like on SF network instances generated with configuration model, we input its empirical degree distribution into the mean-field framework to calculate $n_{g}$.
Results of $K$-core sizes are shown in Fig.\ref{fig:real}.
We can see that, even there are rich structural features in real-world networks beyond the description power of degree distribution, for the network considered here, simulation result and analytical prediction correspond very well.
For the transition behavior, we can find a quite similar picture with that on SF networks and on ER random and RR graphs with $K = 2$.

\section{Result: A cost-benefit analysis}
\label{sec:result_cost}

We present here a cost-benefit analysis of node reinforcement. A random node reinforcement leads to the benefit of an increase in GC of $K$-cores $n_{\rm g}$ at the cost of a fraction $q$ of reinforced nodes. An evaluation taking both the benefit and its cost into consideration can be a better measure of algorithm performance for different node reinforcement schemes.
On a given graph instance $G$ with a node reinforcement fraction $q$ upon an initial fraction $p$ in $K$-core percolation, we define a gain function $g(q; G, p)$ being a benefit term $b(q; G, q)$ minus a cost term $t(q; G)$ as
\begin{eqnarray}
\label{eq:gain_function}
g(q; G, p) = b(q; G, p) - t(q; G).
\end{eqnarray}
Here, $G$ also represents structural parameters of a graph, including degree distribution. The benefit term is defined as the increase of $n_{g}$ due to node reinforcement as
\begin{eqnarray}
\label{eq:benefit_term}
b(q; G, p) = n_{\rm g}(q; G, p) - n_{\rm g}(0; G, p).
\end{eqnarray}
The cost term $t(q, G)$ depends only on reinforced nodes and underlying graph topology.
In a random node reinforcement, we consider it being polynomial of the size of reinforced nodes $q$ as
\begin{eqnarray}
\label{eq:cost_term}
t(q; G) && = w q ^{\alpha},
\end{eqnarray}
in which the coefficient $w (> 0)$ is a relative weight factor between benefit and cost terms, and the dependence parameter $\alpha > 0$.
In the language of economics \cite{Samuelson.Nordhaus-2010-19e}, the cost term with $\alpha < 1$, $\alpha = 1$, and $\alpha > 1$ respectively corresponds to a situation with diminishing, constant, and increasing marginal cost in reinforcing nodes (simply the derivative of cost term with respect to $q$).

We can see that, $g(0; G, p) = 0$.
Based on the above model, given an affordable maximal reinforcement fraction $q_{M} \in (0, 1]$, the optimal choice of random reinforcement corresponds to $q^{\ast}$ with the largest gain $g(q; G, p)$, equivalently
\begin{eqnarray}
\label{eq:optimal_q}
q^{\ast} = {\rm argmax}_{q \in [0, q_{M}]} g(q; G, p).
\end{eqnarray}
Thus, Eqs.(\ref{eq:x}) - (\ref{eq:ng}), (\ref{eq:gain_function}) - (\ref{eq:optimal_q}) establish an analytical approach to locate the optimal fraction in random node reinforcement on uncorrelated random graphs.
For convenience, we set $q_{M} = 1$ in the following discussion.

In Fig.\ref{fig:gain_sublinear}, we show the evolution of $g(q; G, p)$ with $q$ under a cost term with $\alpha = 0.8$ on two random graph models and a real-world network instance.
Its general picture of result is straightforward.
When $w \leqslant w^{(b)}$, the benefit term dominates the gain function, and correspondingly $q^{\ast} = 1$. Yet when $w \geqslant w^{(c)}$, the cost term dominates the gain function, and correspondingly $q^{\ast} = 0$ with $g(q^{\ast}; G, p) = 0$.
Here, $w^{(b)}$ and $w^{(c)}$ are boundary parameters for different regimes of $q^{\ast}$. In most cases of this result, $w^{(b)} = w^{(c)}$. Yet in Fig.\ref{fig:gain_sublinear} (f), when $q = 0.5$ and $0.55$, $q^{\ast} \in (0, 1)$, signifying a nontrivial yet narrow $(w^{(b)}, w^{(c)})$.

In Fig.\ref{fig:gain_linear}, we show the same cases of Fig.\ref{fig:gain_sublinear} with $\alpha = 1$. A quite similar scenario happens, while a clearly discernible $(w^{(b)}, w^{(c)})$ exists in all cases.
For example, in Fig.\ref{fig:gain_linear} (b) for ER random graphs with $K = 3$, 
$w^{(b)} \approx 0.3021$, and $w^{(c)} \approx 0.3725$.
When $w = 0.31 \in (w^{(b)}, w^{(c)})$, the maximal $g = 0.0134535$ when $q^{\ast} \approx 0.606$.

In Fig.\ref{fig:gain_superlinear}, we show the same cases of Fig.\ref{fig:gain_sublinear} with $\alpha = 1.5$.
We can see that $w^{(c)}$ disappears here, equivalently $w^{(c)} \to +\infty$. The reason is that for very small $q$, the cost term is negligible, leading to a positive gain function, thus there is no regime of $q^{\ast} = 0$ in this case. A proper $w^{(b)}$ still exists, and when $w > w^{(b)}$, $q^{\ast} \in (0, 1)$. For example, in Fig.\ref{fig:gain_superlinear} (b) for ER random graphs with $K = 3$,  $w^{(b)} \approx 0.2016$. When $w = 0.3 (> w^{(b)})$, the maximal $g = 0.0621818$ when $q^{\ast} \approx 0.490$.

Summing the main conclusion in Figs.\ref{fig:gain_sublinear} - \ref{fig:gain_superlinear}, for cost terms with linear and superlinear forms, respectively cases with constant and increasing marginal costs, with an intermediate weight factor $w \in (w^{(b)}, w^{(c)})$, the gain function shows a unimodality, and a nontrivial optimal reinforcement fraction $q^{\ast} \in (0, 1)$ can be identified from the maximal gain function with simple numerical methods.

Besides, there are two more observations. The second observation is that, for a given $K$, the ER random graphs, the SF networks, and the real-world network show an increasing weight factor $w$ when we achieve the same maximal gain function.
It partially originates from their increasing heterogeneity of degree profiles and the nontrivial higher-order structure ubiquitous in real-world networks.
The third observation is that, on a given graph instance, when $K$ increases,
the range of large gains around the maximum of gain function becomes smaller,
leading to a narrower region of $q$ in which we can operate at a level of large gains.

We should mention here that, other than reinforcing nodes, reinforcement procedures in different natures can be defined, such as reinforcing links to join non-neighboring functioning clusters into one, or cooperatively reinforcing multiple nodes other than a single node in a reinforcement step. After choosing proper forms for cost functions in Eq.(\ref{eq:gain_function}), a comparison of performances between different network reinforcement schemes and finding the optimal one can be carried out in a systematic and quantitative way.

\section{Conclusion}
\label{sec:conclusion}

In this paper, we consider how a random node reinforcement in a network reshapes its mesoscopic structure through the lens of $K$-cores upon random failures.
Combining our mean-field theory with simulation,
we first show that random node reinforcement increases sizes of $K$-cores,
moves their birth points to a smaller control parameter,
and smoothes a former sudden emergence of $K$-cores gradually into a continuous one.
We then present a cost-benefit analysis for random reinforcement scheme,
and show that given a cost function with a weight factor,
we can numerically find the optimal reinforcement fraction to achieve the largest gain,
thus further make possible comparing various network reinforcement schemes in a single framework.

In all, our framework helps to understand how the methods to enhance network robustness
affect network structure at a refined level and to develop optimal schemes for a targeted approach based on $K$-core structure to build robust artificial systems.

\section*{Acknowledgements}

J.-H. Zhao is supported by Guangdong Major Project of Basic and Applied Basic Research No. 2020B0301030008, Guangdong Basic and Applied Basic Research Foundation (Grant No. 2022A1515011765), and National Natural Science Foundation of China (Grant No. 12171479).
Y. Hu is supported by Natural Science Foundation of Guangdong for Distinguished Youth Scholar, Guangdong Provincial Department of Science and Technology (Grant No. 2020B1515020052), Guangdong High-Level Personnel of Special Support Program, Young TopNotch Talents in Technological Innovation (Grant No. 2019TQ05X138), and National Natural Science Foundation of China (Grant No. 12275118).

\section*{Appendix}

We modify Eqs.(\ref{eq:x}) - (\ref{eq:ng}) in the main text as
\begin{eqnarray}
x
&& = [q + (1 - q) p] - (1 - q) p \sum _{s = 0}^{K - 2} x^{s} Q^{(s)}(1 - x), \\
x_{\rm g}
&& =  [q + (1 - q) p] -  [q + (1 - q) p] Q^{(0)}(1 - x_{\rm g}) \nonumber \\
&& - (1 - q) p \sum _{s = 0}^{K - 2} [x^{s} - (x - x_{\rm g})^{s}] Q^{(s)}(1 - x), \\
n
&& =  [q + (1 - q) p] - (1 - q) p \sum _{s = 0}^{K - 1} x^{s} P^{(s)}(1 - x), \\
n_{\rm g}
&& =  [q + (1 - q) p] - [q + (1 - q) p] P^{(0)}(1 - x_{\rm g}) \nonumber \\
&& - (1 - q) p \sum _{s = 0}^{K - 1} [x^{s} - (x - x_{\rm g})^{s}] P^{(s)}(1 - x).
\end{eqnarray}
In the above equations, we define generating functions for $Q(k)$ and $P(k)$ respectively as
$Q^{(s)}(x) \equiv \sum _{k = s + 1}^{+\infty} Q(k) {k - 1 \choose s} x ^{k - 1 - s}$ and
$P^{(s)}(x) \equiv \sum _{k = s}^{+\infty} P(k) {k \choose s} x^{k - s}$,
with $s \in \{0, 1, \cdots\}$.

\end{document}